\documentstyle[12pt]{article}
\def\1{\mbox{I\hspace{-.15em}1}}

\def\b{\begin{equation}}
\def\e{\end{equation}}
\def\bee{\begin{enumerate}}
\def\eee{\end{enumerate}}

\title{Spectrum of gravitational waves \\in Krein Space Quantization}
\author{M. Mohsenzadeh$^1$\thanks{e-mail:
mohsenzadeh@qom-iau.ac.ir}, A. Sojasi$^{2}$\thanks{e-mail:
sojasi@iaurasht.ac.ir} and E. Yusofi$^{3}$\thanks{e-mail:
e.yusofi@iauamol.ac.ir}}
\date{\today}

\begin{document}

\maketitle {\it \centerline{$^1$ Departman of physics, Qom branch,
Islamic Azad University, Qom,Iran}  \centerline{\it $^{2}$ Departman of physics, Rasht branch,
Islamic Azad University, Rasht,Iran} \centerline{\it $^{3}$ Departman of physics, Ayatollah Amoli branch,
Islamic Azad University, Amol,Iran}}

\begin{abstract}

The main goal of this paper is to derive the primordial power
spectrum for the scalar perturbations generated as a result of
quantum fluctuations during an inflationary period by an
alternative approach of field quantization
\cite{gareta,ta2002,rota2006}. Formulae are derived for the
gravitational waves, special cases of which include power law
inflation and inflation in the slow roll approximation, in Krein
space quantization.

\end{abstract}


\vspace{0.5cm} {\it Proposed PACS numbers}: 04.62.+v, 03.70+k, 98.80.-k, 04.30.-w \vspace{0.5cm}


\section{Introduction}

In this paper we derive formula for the spectrum of gravitational
waves produced during inflation, for inflation in general in Krein
space quantization. The standard results to first order in the
slow roll approximation in Hilbert space quantization are \b
P_{\psi}^{1/2}=\frac{H}{2\pi}|_{aH=k} \e for the gravitational
wave spectrum\cite{ta}. Thus, the outline of the paper is as
follows. In section $2$ we briefly recall definition of power spectrum of gravitational waves. Section
$3$ is devoted to calculation of power spectrum in Krein space
quantization. Section $4$ is devoted to calculation of power
spectrum in special cases. Brief conclusion and outlook are given
in final section.

\section{Definition of power spectrum of gravitational waves}
Our units are such that $ c=\hbar=8\pi G=1 $. $ H $ is the Hubble
parameter, $ \phi $ is the inflaton field and a dot denotes the
derivative with respect to time $ t $. The background metric
is\cite{gagata} \b
ds^{2}=dt^{2}-a^{2}(t)d{\mathbf{x}}^{2}\\=a^{2}(\eta)(d\eta^{2}-d{\mathbf{x}}^{2})\e
Tensor linear perturbations to (2) can be expressed most generally
as\cite{garota} \b
ds^{2}=a^{2}(\eta)[d\eta^{2}-(\delta_{ij}+2h_{ij})dx^{i}dx^{j}]\e
The spectrum of gravitational waves is defined by\cite{al,mi} \b
h_{ij}=\int\frac{d^{3}{\mathbf{k}}}{(2\pi)^{3/2}}\sum^{2}_{\lambda{=1}}\psi_{\mathbf{k},\lambda}(\eta)\xi_{ij}({\mathbf{k}},\lambda)e^{i{\mathbf{k.x}}}\e
\b
\langle{0}|\psi_{\mathbf{k},\lambda}\psi_{\mathbf{l},\lambda}^{*}|{0}\rangle=\frac{2\pi^{2}}{k^{3}}P_{\psi}\delta^{3}\mathbf(k-l)\e
where $ \xi_{ij}(\mathbf{k},\lambda) $ is a polarization tensor
satisfying\b \xi_{ij}=\xi_{ji}, \xi_{ii}=0, k_{i}\xi_{ij}=0\e
\b
\xi_{ij}({\mathbf{k}},\lambda)\xi_{ij}^{*}(\mathbf{k},\mu)=\delta_{\lambda\mu}\e
It is also useful to
choose\b \xi_{ij}(-{\mathbf{k}},\lambda)=\xi_{ij}^{*}(\mathbf{k},\lambda)
\e
\section{Calculation in Krein space quantization}
In this section we calculate the power spectrum of gravitational
waves in Krein space quantization. First, we briefly
recall the Krein space quantization. In the previous paper
\cite{ta5}, we present the free field operator in the Krein
space quantization. The field operator in Krein space is built by
joining two possible solutions of field equations, positive and
negative norms \b
\phi(\eta,{\mathbf{x}})=\phi_p(\eta,{\mathbf{x}})+\phi_n(\eta,{\mathbf{x}}),\e
where
$$ \phi_p(\eta,{\mathbf{x}})=(2\pi)^{-3/2}\int d^{3}{\mathbf{k}}
[c({\mathbf{k}})u_{k,p}e^{i\textbf{k.x}}+c^{\dag}({\mathbf{
k}})u_{k,p}^*e^{-i\textbf{k.x}}],$$ $$
\phi_n({\eta,\mathbf{x}})=(2\pi)^{-3/2}\int d^{3}{\mathbf{k}}
[b({\mathbf{
k}})u_{k,n}e^{-i\textbf{k.x}}+b^{\dag}({\mathbf{k}})u_{k,n}^*e^{i\textbf{k.x}}],$$
and $c({\mathbf{k}})$ and $b({\mathbf{k}})$ are two independent
operators. Creation and annihilation operators are constrained to
obey the following commutation rules \b [c(
{\mathbf{k}}),c({\mathbf{k'}})]=0,\;\;[c^{\dag}({\mathbf{ k}}),
c^{\dag}({\mathbf{k'}})]=0,\;\;, [c({\mathbf{
k}}),c^{\dag}({\mathbf{k'}})]=\delta({\mathbf{k-k'}}) ,\e \b
[b({\mathbf{k}}),b({\mathbf{ k'}})]=0,\;\;[b^{\dag}({\mathbf{
k}}), b^{\dag}( {\mathbf{k'}})]=0,\;\;, [b({\mathbf{
k}}),b^{\dag}( {\mathbf{k'}})]=-\delta({\mathbf{ k-k'}}) ,\e
 \b [c({ \mathbf{k}}),b({\mathbf{k'}})]=0,\;\;[c^{\dag}({\mathbf{k}}), b^{\dag}({\mathbf{ k'}})]=0,\;\;, [c
({\mathbf{k}}),b^{\dag}({\mathbf{k'}})]=0,\;\;[c^{\dag}(
{\mathbf{k}}),b({\mathbf{k'}})]=0 .\e The vacuum state $\mid
\Omega>$ is then defined by \b c^{\dag}({\mathbf{k}})\mid \Omega>=
\mid 1_{{\mathbf{ k}}}>;\;\;c({\mathbf{k}})\mid \Omega>=0, \e \b
b^{\dag}({\mathbf{ k}})\mid \Omega>= \mid
\bar1_{{\mathbf{k}}}>;\;\;b({\mathbf{ k}})\mid \Omega>=0, \e \b
b({\mathbf{ k}})\mid 1_{ {\mathbf{k}}}
>=0;\;\; c({\mathbf{k}})\mid \bar1_{{\mathbf{k}}} >=0, \e
where $\mid 1_{{\mathbf{k}}}> $ is called a one particle state and
$\mid \bar1_{{\mathbf{k}}}> $ is called a one ``unparticle
state''. By imposing the physical interaction on the field
operator, only the positive norm states are affected. The negative
modes do not interact with the physical states or real physical
world, thus they can not be affected by the physical interaction
as well.

The action for tensor linear
perturbation is\cite{ta2001} \b S=\frac{1}{2}\int
a^{2}[(h_{ij}')^{2}-(\partial_{l}h_{ij}^{2})]d\eta
d^{3}{\mathbf{x}} =\frac{1}{2}\int d^{3}k  \times
\sum_{\lambda{=1}}^{2}\int
[|\upsilon'_{{\mathbf{k}},\lambda}|^{2}-(k^{2}-\frac{a''}{a})|\upsilon
_{{\mathbf{k}},\lambda}|^{2}]d\eta\e where \b \upsilon
_{{\mathbf{k}},\lambda}=a\psi_{{\mathbf{k}},\lambda}\e[N.B.$
\upsilon_{{\mathbf{k}},\lambda}=\upsilon^{*}_{{-\mathbf{k}},\lambda}
 $
from (4) and (8)]. Quantizing \b
\upsilon_{{\mathbf{k}},\lambda}(\eta)=\{\upsilon_{k,p}(\eta)c({\mathbf{k}},\lambda)+\upsilon^{*}_{k,p}(\eta)c^{\dag}({-\mathbf{k}},\lambda)\}+
\{\upsilon_{k,n}(\eta)b({\mathbf{k}},\lambda)+\upsilon^{*}_{k,n}(\eta)b^{\dag}({-\mathbf{k}},\lambda)\}\e

The equation of motion for $ \upsilon_{k} $ is \b
\upsilon''_{{\mathbf{k}}}+(k^{2}-\frac{a''}{a})\upsilon_{{\mathbf{k}}}=0\e
and \b \upsilon_{k}(\eta)\sim \frac{1}{\sqrt{2k}}e^{\pm ik\eta},\;\;
 as\;\;\frac{k}{aH}\gg {1} \e
\b \upsilon_{k}\sim a ,\;\; as\;\;\frac{k}{aH}\ll {1} \e Assuming $ \epsilon\equiv-\frac{\dot{H}}{H^{2}} $ is constant, and \b
\frac{a''}{a}=2a^{2}H^{2}(1-\frac{1}{2}\epsilon)=\frac{1}{\eta^{2}}(\mu^{2}-\frac{1}{4})\e
where $ \mu=\frac{1}{1-\epsilon}+\frac{1}{2} $. Now \b \langle
0|\psi_{{\mathbf{k}},\lambda}
\psi^{*}_{{\mathbf{l}},\sigma}|0\rangle=\frac{1}{a^{2}}(|\upsilon_{k,p}|^{2}+|\upsilon_{k,n}|^{2})\delta_{\lambda
\sigma}\delta^{3}({\mathbf{k-l}})\e Therefore from (5) and (23)
\b
P_{\psi}(k)=\frac{k^{3}}{2\pi^{2}}\frac{|\upsilon_{k,p}(\eta)|^{2}-|\upsilon_{k,n}(\eta)|^{2}}{a^{2}}\e

\section{Special cases}

\subsection{slow-roll approximation}

The orthogonal eigenmodes $ u_{k} $, $ u^{*}_{k} $ of (19) are
easy to construct in the slow roll approximation, when $ \epsilon
$ and $ \mu $ are small,$ \epsilon, \mu \ll {1} $. During
slow roll inflation, to order in $ \epsilon $ and $\mu $, we have $
(1-\epsilon)\eta=-\frac{1}{aH} $ and so $
\frac{a''}{a}=\frac{2-3\mu+6\epsilon}{\eta^{2}} $. Then the
mode equation (19) become\cite{mi,ta3} \b
u''_{k}+(k^{2}-\frac{2-3\eta+6\epsilon}{\eta^{2}})u_{k}=0\e

The standard choice\cite{ta2} of the eigenmodes $ u_{k},u^{*}_{k}
$ is to take \b
u_{k}(\eta)=-\frac{\sqrt{\pi\eta}}{2}H^{(-)}_{\nu}(k\eta),u^{*}_{k}(\eta)=-\frac{\sqrt{\pi\eta}}{2}H^{(+)}_{\nu}(k\eta)
\e as the positive and negative frequency modes, respectively,
where $ \nu=\frac{3}{2}-\mu+2\epsilon $\cite{mi}. The
normalization of $ u_{k} $ is chosen such that eqs.(10)-(15)
follows from the canonical commutation relations $
[u(\eta,{\mathbf{x}}),\Pi(\eta,{\mathbf{x'}})]=i\delta^{3}({\mathbf{x-x'}})
$. Using $ |0\rangle $ as the state of the inflaton during
inflation and ignoring the slow roll corrections, in which case
the eigenmodes (26) reduce to \b
u_{k,p}=\frac{1}{\sqrt{2k}}(1-\frac{i}{k\eta})e^{-ik\eta},u_{k,n}=\frac{1}{\sqrt{2k}}(1+\frac{i}{k\eta})e^{ik\eta}\e
Then, according to the second perspective of \cite {ta5} \b
\langle{0}|\psi_{\mathbf{k,\lambda}}\psi_{\mathbf{l,\lambda}}^{*}|{0}\rangle=\frac{H}{2k^{3}}(ke^{-\alpha
k^{2}})\delta^{3}\mathbf(k-l)\e
Therefore we have for power spectrum from (5) and (28) \b P_{\psi}(k)=\frac{H}{4\pi^{2}}ke^{-\alpha k^{2}}\e
where $ \alpha=\frac{1}{\pi H^{2}}
$ , is related to the density of
gravitons\cite{rota2006,rota1}.

\subsection{Power-law inflation }

In power law inflation for the mode equation (19) we need $ a''/a
$. To compute this we have \b a(t)\propto t^{p}\e Then $
t\propto \eta^{1/1-p} $ , So $ a(\eta)\propto \eta^{p/1-p} $.
Hence,\b \frac{a''}{a}=(\nu^{2}-\frac{1}{4})\frac{1}{\eta^{2}}\e
where \b \nu^{2}-\frac{1}{4}=\frac{p(2p-1)}{(p-1)^{2}}\e

Using this in (19) gives the mode equation\b
u''_{k}+(k^{2}-\frac{\nu^{2}-1/4}{\eta^{2}})u_{k}=0\e This can be
solved in terms of Bessel functions. Before proceeding with this
we note two further relations. First, from $ H=\frac{p}{t} $ and $
a(t)=a_{\circ}t^{p} $ we get \b
\eta=-\frac{1}{aH}\frac{1}{1-1/p}\e In addition \b
\epsilon=\frac{1}{p}=constant\e

Let us now turn to the mode equation (33). According to \cite{ja},
the functions $ \omega(z)=z^{1/2}c_{\nu}(\lambda z) $, $
c_{\nu}\propto H^{(1)}_{\nu}, H^{(2)}_{\nu},... $ satisfy the
differential equation \b
\omega''+(\lambda-\frac{\nu^{2}-1/4}{z^{2}})\omega=0\e From the
asymptotic formula for large $ k\eta $ \cite{ja} \b
H_{\nu}(k\eta)\sim \sqrt{\frac{2}{\pi
k\eta}}[1-i\frac{4\nu^{2}-1}{8k\eta}]exp[-ik\eta-i\pi(\frac{\nu}{2}+\frac{1}{4})]
\e According to the second perspective of \cite{ta5}, we see that the correct solutions are  \b
\upsilon_{k,p}(\eta)=\frac{1}{\sqrt{2k}}(1-\frac{i}{k\eta})e^{i(\mu+\frac{1}{2})\frac{\pi}{2}}e^{-ik\eta}\e
\b
\upsilon_{k,n}(\eta)=\frac{1}{\sqrt{2k}}(1+\frac{i}{k\eta})e^{-i(\mu+\frac{1}{2})\frac{\pi}{2}}e^{ik\eta}\e
where $ \mu $ \b \mu=\frac{3}{2}+\frac{1}{p-1}\e  Therefore from (38),(39) and using (5),(23) we have: \b P_{\psi}(k)=\frac{H}{4\pi^{2}}ke^{-\alpha k^{2}}\e

\section{Conclusion}

The negative frequency solutions of the field equations are needed
for the covariant quantization in the minimally coupled scalar
fields in de Sitter space. Contrary to the Minkowski space, the
elimination of de Sitter negative norms in this case breaks the de
Sitter invariance. In other words, in order to restore the de
Sitter invariance, one needs to take into account the negative
norm states {\it i.e.} the Krein space  quantization. This
provides a natural tool for renormalization of the theory
\cite{gareta}. The spectrum of gravitational waves, has been calculated through the Krein space
quantization exhibiting. Once again the theory is automatically
renormalized.\vspace{0.5cm}

\noindent {\bf{Acknowlegements}}:This work has been supported by the Islamic Azad
University-Qom Branch, Qom, Iran.

\end{document}